\documentclass[%
 reprint,
superscriptaddress,
frontmatterverbose,
 amsmath,amssymb,
 APS,
]{revtex4-2}

\usepackage{graphicx}
\usepackage{dcolumn}
\usepackage{bm}
\usepackage{color}
\usepackage{siunitx}
\preprint{APS/123-QED}

\begin{document}

\title{Magnetic electron collimation in three-dimensional semi-metals}

\author{Xiangwei Huang}
\affiliation{Laboratory of Quantum Materials (QMAT), Institute of Materials (IMX), \'Ecole Polytechnique F\'ed\'erale de Lausanne (EPFL), 1015 Lausanne, Switzerland}

\author{Carsten Putzke}%
\affiliation{Laboratory of Quantum Materials (QMAT), Institute of Materials (IMX), \'Ecole Polytechnique F\'ed\'erale de Lausanne (EPFL), 1015 Lausanne, Switzerland}

\author{Chunyu Guo}%
\affiliation{Laboratory of Quantum Materials (QMAT), Institute of Materials (IMX), \'Ecole Polytechnique F\'ed\'erale de Lausanne (EPFL), 1015 Lausanne, Switzerland}

\author{Jonas Diaz}%
\affiliation{Laboratory of Quantum Materials (QMAT), Institute of Materials (IMX), \'Ecole Polytechnique F\'ed\'erale de Lausanne (EPFL), 1015 Lausanne, Switzerland}

\author{Markus K\"onig}
\affiliation{Max Planck Institute for Chemical Physics of Solids, 01187 Dresden, Germany}%

\author{Horst Borrmann}
\affiliation{Max Planck Institute for Chemical Physics of Solids, 01187 Dresden, Germany}%

\author{Nityan L. Nair}
\affiliation{Department of Physics, University of California, Berkeley, 94720 California, USA}

\author{James G. Analytis}
\affiliation{Department of Physics, University of California, Berkeley, 94720 California, USA}

\author{Philip J.W. Moll}
\email[Email:] {Philip.Moll@epfl.ch}
\affiliation{Laboratory of Quantum Materials (QMAT), Institute of Materials (IMX), \'Ecole Polytechnique F\'ed\'erale de Lausanne (EPFL), 1015 Lausanne, Switzerland}




\date{\today}

\begin{abstract}

While electrons moving perpendicular to a magnetic field are confined to cyclotron orbits, they can move freely parallel to the field. This simple fact leads to complex current flow in clean, low carrier density semi-metals, such as long-ranged current jets forming along the magnetic field when currents pass through point-like constrictions. Occurring accidentally at imperfect current injection contacts, the phenomenon of ``current jetting" plagues the research of longitudinal magneto-resistance which is particularly important in topological conductors. Here we demonstrate the controlled generation of tightly focused electron beams in a new class of micro-devices machined from crystals of the Dirac semi-metal Cd$_3$As$_2$. The current beams can be guided by tilting a magnetic field and their range tuned by the field strength. Finite element simulations quantitatively capture the voltage induced at faraway contacts when the beams are steered towards them, supporting the picture of controlled electron jets. These experiments demonstrate the first direct control over the highly non-local signal propagation unique to 3D semi-metals in the current jetting regime, and may lead to novel applications akin to electron optics in free space.

\end{abstract}

\maketitle

Generating and controlling electron beams by electric and magnetic fields has been a highly successful concept in technology, such as cathode-ray tubes or electron microscopes. Translating these ideas from electron optics to solid state devices with new functionalities is a long-standing research challenge\cite{vanHouten1995}. Unperturbed free-space-like electron motion in a metal requires long mean-free-paths, and hence clean materials with high carrier mobility. In such good conductors, however, external electric fields are effectively screened and thus cannot be used to shape the electronic motion. Interesting results to generate beams rely on geometric shaping of constrictions in the electron pathway in ballistic devices, akin to shaping a light beam by a pinhole in the geometric optics limit. Two dimensional electron gases (2DEG) are ideally suited to pursue these ideas\cite{beenakker1991quantum,bhandari2018imaging,sivan1990electrostatic,spector1990electron,cheianov2007focusing,hartmann2010smooth,williams2011gate,spector1990refractive,chen2016electron,van1989coherent}, as they naturally offer the long mean-free-paths and the high level of geometric control required to shape complex pinhole constrictions\cite{barnard2017absorptive,rickhaus2015gate}. At the same time, restricting the electrons to two dimensions reduces the richness of the equations of motion, and thereby severely limits the mechanisms to generate beams. For example, the orbital motion of electrons in 2DEGs is entirely determined by the out-of-plane component of the magnetic field, thus reducing the magnetic field effectively to a scalar quantity. Further, due to the need of a restriction, the beam shape is set at the time of fabrication. Both remove the possibility to shape and steer beams by changing the direction of the field as used in free space.

The main goal of this work is to experimentally demonstrate electron beam control in devices made from three-dimensional semi-metals and highlight the additional degrees of freedom for electronic motion. Unlike the two-dimensional case, in three dimensions in-plane magnetic fields induce an out-of-plane Hall component that significantly reshapes the current flow. This can be visualized by the helical trajectories of electrons entering a bulk metal isotropically through a point-like contact (Fig.1a). As the Lorentz force only acts on the particle velocity component perpendicular to the magnetic field, the effective conductivity describing the metal must be anisotropic even in conductors which are isotropic in zero field. 

These features can be captured by a conductivity tensor $\sigma_{\rm ij}$, relating current $J$ and electric field $E$ via Ohms law, $J_{\rm i}\,=\,\sigma_{\rm ij}E_{\rm j}$. Assuming an isotropic metal in zero field, $\sigma_{\rm xx}=\sigma_{\rm yy}=\sigma_{\rm zz}=\sigma_0$, the magnetic field $B$ applied along the $\mathbf{z}$-direction induces a transport anisotropy:

$
\sigma=\begin{pmatrix}
\sigma _{\rm xx}(B) &\sigma _{\rm xy}(B) & 0\\
-\sigma _{\rm xy}(B) & \sigma _{\rm xx}(B) & 0\\
0&0  & \sigma _0
\end{pmatrix}$.

While the material-dependent details of the Fermi surface geometry and scattering matrix determine the conductivity, the main features discussed here are captured by a simple Drude picture\cite{pippard2009magnetoresistance}, $\frac{\sigma _{\rm xx}(B)}{\sigma _0}= \frac{1}{1+(\mu B)^{2}}$ . In this approximation, the conductivity parallel to the magnetic field $\sigma _{\rm zz}=\sigma_0$ remains unchanged, while the orthogonal electronic motion is strongly suppressed due to the Lorentz force. When $\mu B \gg 1$ and hence $\sigma _{\rm zz}\gg\sigma _{\rm xx}$, the magnetic field induces a significant current redistribution and enforces a preferential flow along the magnetic field direction. As we will find later, the large Hall conductivity of low carrier density materials is of essence for our goal to obtain long range current beams.

While this generic consideration well describes the situation in many semi-metals, here we focus experimentally on the Dirac semi-metal Cd$_{3}$As$_{2}$. The conductivity along different crystal directions shown in Fig.1b is measured on a multi-terminal device as shown in Fig.5a. The appearance of quantum oscillations at low fields ($B\,\sim\,1\,$T) and the low oscillation frequency ($F\,\sim\,27\,$T) evidence its high mobility $\mu$ and low carrier density $n$. A complete analysis based on the Drude model and Hall resistivity (see Supplementary Figure 4 ) yields $\mu\,\sim 1.84\,\rm m^{2}\rm V^{-1}\rm s^{-1}$ and $n\,=\,6.54\times10^{24}\,\rm m^{-3}$ at 2\,K, in agreement with other reports on this compound\cite{Liang2015,Zhao2015}. Accordingly, its magneto-conductivity when the current is parallel ($\sigma_{\parallel}$) and perpendicular ($\sigma_{\perp}$) to the magnetic field $B$ is strikingly different even at low magnetic field (Fig.1b). The anisotropy $A\,=\,\sigma_{\parallel}/\sigma_{\perp}$ rapidly increases with magnetic field and reaches a value of 706 at 14\,T. Therefore, strong current path reorientations would be expected in this compound.

Yet strong magneto-resistance anisotropy alone does not induce any transport anomalies. Key to the appearance of long-ranged, non-uniform current patterns in magnetic fields are constrictions in the current flow as they geometrically enforce electron velocity components parallel and perpendicular to the magnetic field. This situation and the emergence of current beams can be visualized by solving Laplace equations in the presence of point-like current sources placed in close proximity on the top of a cuboid structure. Finite element calculations (see methods) of the electric potential are given in Fig.1d-f for three different anisotropies $A$\,=\,1,\,10,\,706, corresponding to the experimentally observed situations (Fig.1c). In zero field (Fig.1d), the current flow is confined closely to the vicinity of the contacts. As the magnetic field and hence the conductivity anisotropy increases, the current path changes into elongated current beams ranging deep into the sample. This classical magneto-transport phenomenon has been identified early-on as a main source of erroneous measurements of the longitudinal magneto-resistance. It was shown that imperfections in electric contacts, transmitting currents into a sample only at a few points, lead to non-local plumes of current, known as `current jetting'\cite{yoshida1975anomalous,yoshida1976geometrical}. This effect can cause problems for the interpretation of voltages in a four-probe configuration in terms of resistivities when the field and current are parallel\cite{yoshida1979transport,yoshida1979transport2}. 

Current jetting has plagued already the early research of semi-metals\cite{pippard2009magnetoresistance}, and recently shifted into focus again due to the predictions of negative magneto-resistance in topological semi-metals for parallel magnetic and electric fields\cite{son2013chiral}. As topological semi-metals such as Cd$_3$As$_2$ are generally high mobility, low carrier density materials\cite{zhang2016signatures,li2016chiral}, they are prone to this effect which has been argued to well capture apparent negative longitudinal magneto-resistance in some cases simply due to magnetic field induced current path reorientation\cite{dos2016search}. Similarly, other exotic magneto-transport phenomena expected in topological materials such as the planar Hall effect may also be influenced by current jetting\cite{yang2019current}. Current jetting is generally considered a nuisance due to uncontrolled, point-like current injection and elaborate experiments based on multiple measurements of the electric potential have been designed to eliminate its influence from measurements\cite{liang2018experimental}.

Here we explore the opposite approach to exploit current jetting as a mean to generate and control unconventional current paths in solids. These ideas are conceptually related to longitudinal electron focusing (LEF) experiments performed on macroscopic crystals of semi-metals in a magnetic field. Using a tightly focused laser, the propagation of the thermally excited quasi-particles through the crystal emerging from the focal point could be mapped out\cite{primke1997imaging}. In our approach, we use Focused Ion Beam (FIB) machining of high quality single crystals\cite{moll2016transport} to carve artificial constrictions on the \SI{}{\micro\metre} scale into the current flow. Thereby, current jetting can be induced on purpose and in a controlled way, resulting in tightly focused electron beams which can be guided in the crystal by the magnetic field.

\begin{figure}
	\centering
	\includegraphics[width = 1\linewidth]{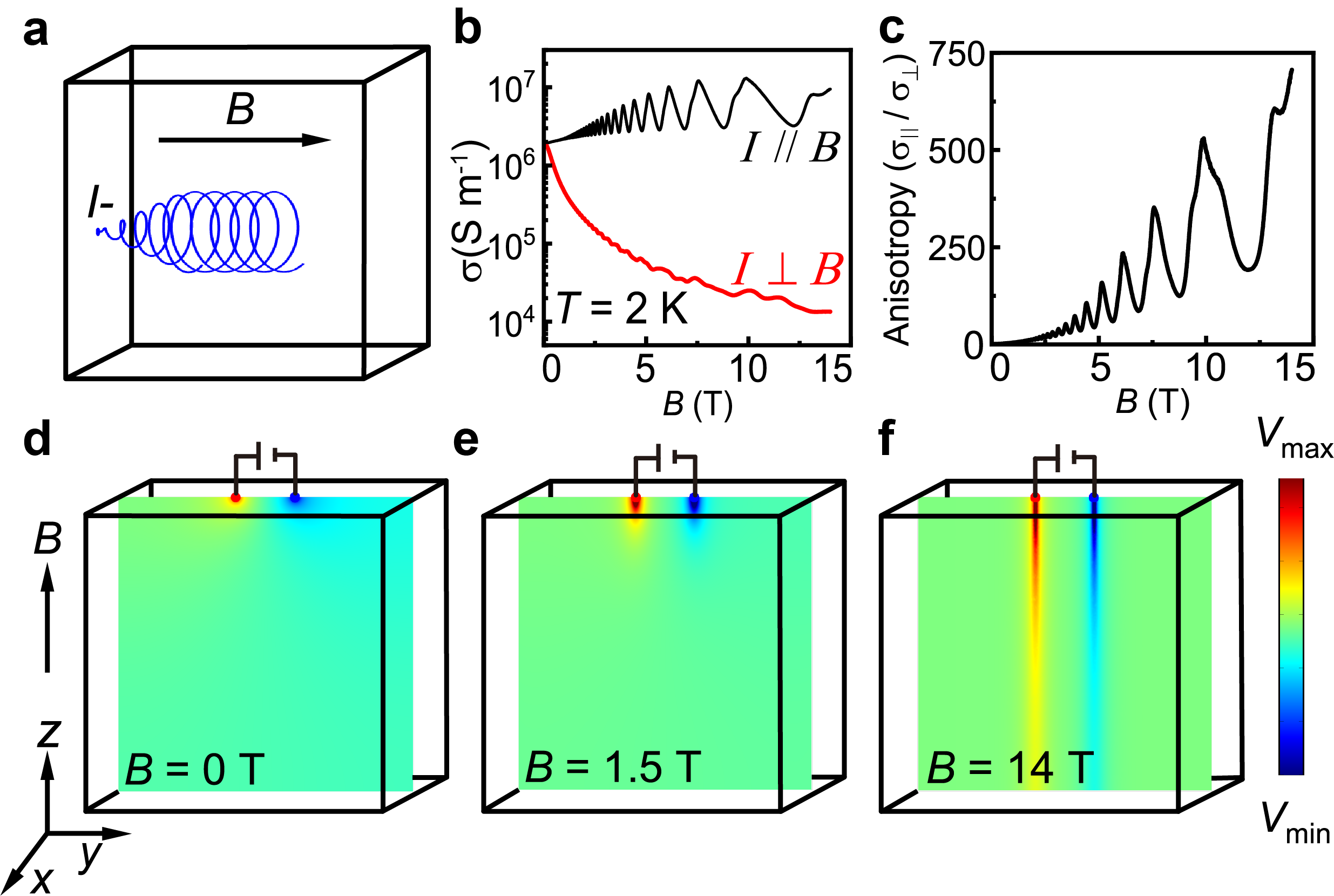}
	\caption{\textbf{a}, Sketch of electron trajectory entering the sample through a point-like contact. \textbf{b}, Magneto-conductivity $\sigma(B)$ of Cd$_{3}$As$_{2}$ when the magnetic field perpendicular $\sigma_{\perp}$ and parallel $\sigma_{\parallel}$ to the current direction, and \textbf{c} their resulting anisotropy $A =\sigma_{\parallel}/\sigma_{\perp}$.  \textbf{d}-\textbf{f}, Calculated electric potential distribution of a 3D square sample with two point current injections at three different magnetic fields applied along the $z$-direction. The color scale is different in the three panels for clarity.}
	\label{Fig1}
\end{figure}



A typical micro-structure carved from flux-grown single crystals of Cd$_3$As$_2$ consists of a main cuboid part connected through four around \SI{1}{\micro\metre} wide crystal bridges which serve as artificial constrictions (Fig.2a, see methods). It is important to note that these are very thick, bulk-like structures far from the 2D limit to ensure fully three-dimensional electron motion. In this case, the sample thickness $t$\,=\,\SI{1.4}{\micro\metre} is much larger than the Fermi wavelength $\lambda_F \sim 17\,$nm \cite{Liang2015}, hence no quantization or sub-band formation is expected in the device along any direction and true 3D behavior should appear, which is self-consistently evidenced by the observation of bulk-like quantum oscillations in the devices. Electric contacts are formed to the outer parts of the structure beyond the constriction by gold deposition and the device itself is supported by a layer of epoxy glue on a sapphire chip. Here, a nozzle spacing of $d$\,=\,\SI{2.0}{\micro\metre} and an almost square main body of $L$\,=\,\SI{11.0}{\micro\metre} was chosen, yet these parameters were varied during this study (see Supplementary Table 1). A current is applied to one pair of nozzles, while the voltage difference between the others is recorded. In zero field, no voltage is detected on the faraway contact pair, as expected for isotropic conductors.

\begin{figure}
	\centering
	\includegraphics[width = 1\linewidth]{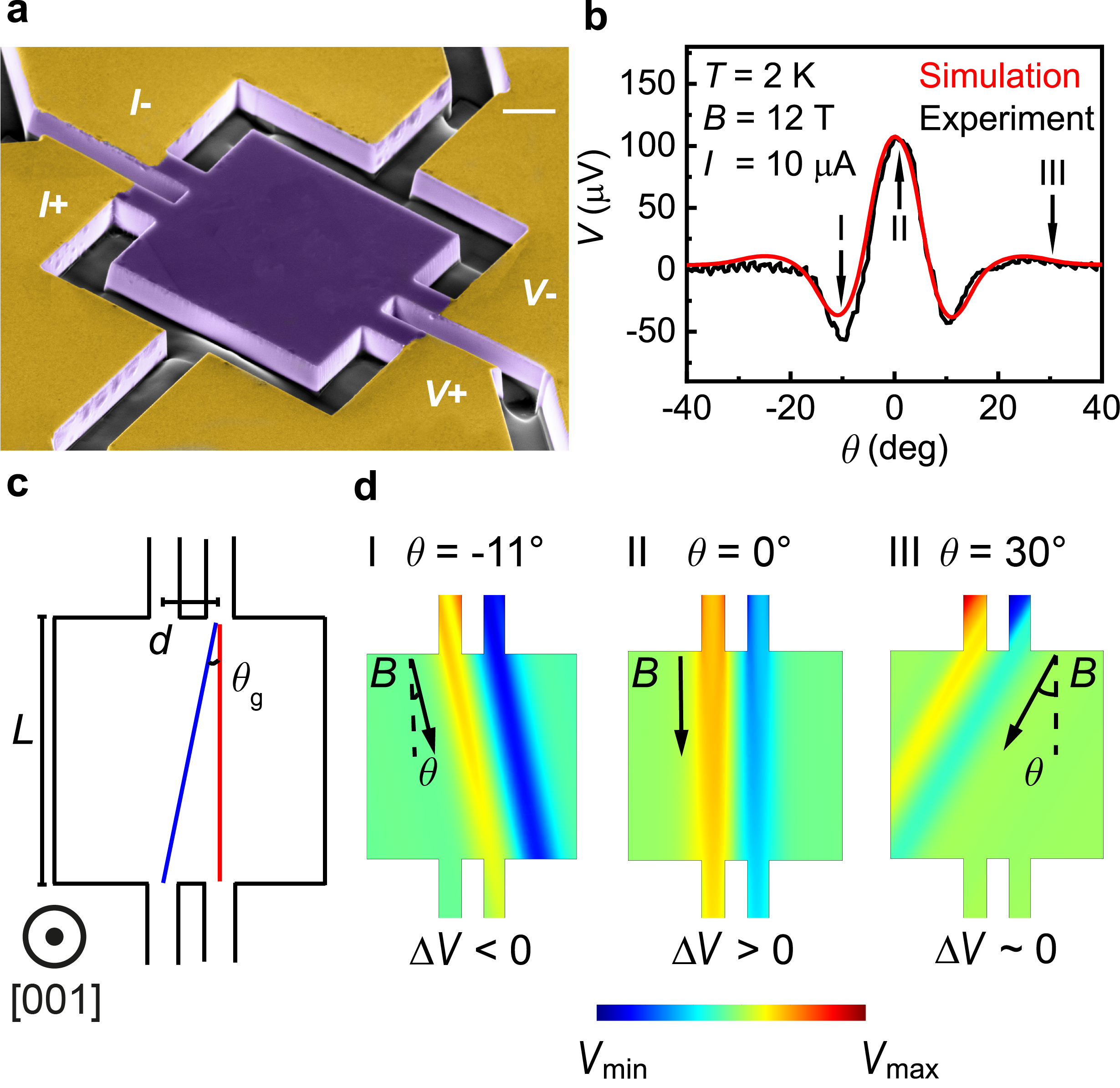}
	\caption{\textbf{a}, False-colour electron micro-graph of a Cd$_{3}$As$_{2}$ (purple) device with Au contacts (yellow)  after micro-structuring in the FIB. The scale bar is \SI{2.0}{\micro\metre} in width. The current is sourced from the upper two contacts, and the voltage is measured from the two bottom electrodes. \textbf{b}, Measured and simulated signals depend on the magnetic field angle $\theta$ (2\,K, 12\,T). The three arrows mark pronounced features: (I) a minimum around $-11^{\circ}$, (II) a strong maximum at $0 ^{\circ}$ and (III) negligible voltage at $30^{\circ}$. \textbf{c}, Schematic of the device in \textbf{a}: $d$\,=\,\SI{2.0}{\micro\metre}, $L$\,=\,\SI{10.7}{\micro\metre}, $\theta_{\rm g}$\,=\,10.6$^{\circ}$. [001] axis of the device is in the out-of-plane direction. \textbf{d}, Electric potential simulation of the device at three positions marked by the arrows in (b). The color scale is offset between panels for clarity.}
	\label{Fig2}
\end{figure}

Applying a magnetic field in the plane of the device can induce a strong voltage in the opposite contact pair (Fig.2). The nature of this non-local transport is ideally elucidated by considering a rotation of a constant magnetic field ($B\,=\,12\,$T) in the sample plane uncovering a characteristic W-shape as a function of field angle including regions of inverted voltage. A large, positive voltage is observed for fields applied along the direction connecting the current and voltage contacts ($\theta\,=\,0^\circ$). Rotating the field in the plane quickly reduces the voltage, leading to a sign change and a pronounced minimum at a small angle around $\theta\,\sim\, 11^\circ$ in this device. At larger angles, the signal increases again until it vanishes. 

This characteristic signal can be directly understood from the generation of current beams following the magnetic field direction (Fig.2d). As the conductivity perpendicular to the magnetic field is low, the current spreads in a beam-like fashion throughout the sample. At $\theta\,=\,0^\circ$, both beams emitted from the nozzles hit their receiving counterparts, leading to a large and positive voltage difference reflecting the positive potential between the current electrodes required to source the current. As the field angle steers the beams away from the center, this situation is reversed and the beam of opposite polarity hits the diagonal contact thus causing the sign change. Indeed, this geometric condition of a straight beam hitting the diagonal electrode, tan($\theta_{\rm g}$)\,=\,$\frac{d}{L}$, quantitatively matches the angle of the voltage minimum. A total of 6 devices of different size $L$ and nozzle separation $d$ yield highly consistent results and the minimum position traces well their respective angle $\theta_{\rm g}$ (see Supplementary Figure 2). At even higher angles, both beams miss the electrodes thus no voltage is observed.

This intuitive picture is quantitatively supported by finite element simulations solving the Laplace equation in the Drude approximation for the sample geometry taking the finite width of the constrictions into account (Fig.2). Here, all parameters used in the calculation, $\sigma_0\,=\,1.35\times 10^{6}$\,S m$^{-1}$ and $\mu B\,=\,36$, were determined from the zero-field resistivity and quantum oscillations of the square device respectively (see Supplementary Figure 3). This simplistic model captures the entire angle dependence of the voltage signal, thus strongly supporting the picture of field-guided current beams. The simulated voltage has been scaled by a factor of 3 onto the experimental data in Fig.2b. This amplitude difference may be removed by using the Drude parameters as fitting parameters, yet here we keep them fixed at the independently determined values. It is interesting to note that both experiment and simulations show the absolute value of the voltage minimum at half the value of the maximum, as expected for a one-beam hit compared to a two-beam hit. This further supports the notion that control over current jetting can be successfully achieved in such devices.


Next we discuss the temperature and magnetic field strength dependence of the electron beams (Fig.3). Increasing the temperature reduces the signal, yet the W-shape is well observed above room temperature and its amplitude at 350\,K remains at 20$\,\%$ of that at 2\,K for field of 14\,T. Similarly, decreasing the magnetic field reduces the signal, which at low temperatures is still well visible at 1\,T. To quantify the strength of current collimation, we estimate the size of the non-local anomaly as the difference between the minimal and maximal voltage, $\Delta=V_{\rm max}-V_{\rm min}$. This antagonistic behavior of temperature and field can be quantitatively understood as the field-induced anisotropy is tuned by the parameter $\mu(T) B$. The loss in mobility, $\mu(T)$, upon increasing temperature can be compensated by an according increase in the magnetic field. In addition, the temperature sets the zero magnetic field conductivity $\sigma_0(T)$ leading to deviations from purely $\mu(T) B$ scaling. Remarkably, the simplistic and fitting-parameter-free Drude approximation describes the device response well over a wide range of temperatures and magnetic fields. Small deviations are to be expected due to finite size corrections, the peculiar field-dependence of the magneto-resistance in Cd$_3$As$_2$\cite{Narayanan2015}, and the oversimplifying assumption of a temperature-independent carrier density.

In this initial study we demonstrate the feasibility of microscopic control over current beams in Cd$_3$As$_2$, yet the field scales required to operate a device from this material at room temperature are impractical. However, any gain in mobility allows to reduce the magnetic field by the same amount, and 3D materials with higher mobilities can be engineered and optimized. An interesting route may be elemental semi-metals such as Bi or Sb. These can be deposited as thick layers and reach very high mobilities via annealing\cite{shoenberg1939magnetic,oktu1967galvanomagnetic}. Further natural candidates are other topological semi-metals, as they combine high Fermi velocities with high mobilities and low carrier densities. The here presented quantitative control over semi-classical transport presents a step towards understanding transport processes arising from the topologically non-trivial nature of Cd$_3$As$_2$ which would appear as discrepancies not explained by semi-classics. One example is the negative longitudinal magneto-resistance due to the chiral anomaly, observed in Cd$_3$As$_2$ nano-conductors\cite{li2015giant}. One direct consequence of this work in the field of topology concerns the recent proposals of field-induced non-local transport due to the Weyl-orbits in topological semi-metals\cite{baum2015current}. As we demonstrate even the classical magneto-transport in such semi-metals to be of a pronounced non-local character without invoking topology, it will be important to formulate a complete theory that self-consistently takes the semi-classical and topological non-locality into account. In addition, transport in topological surface states has been shown to contribute to the transport in micro- and nano-structures of Cd$_3$As$_2$\cite{moll2016transport,zhang2017evolution}. In this work, the crystals were aligned such that the internode direction ($\mathbf{c}$-direction) is perpendicular to the main face of the device. Hence no Fermi arcs are expected to exist on them and the transport is predominantly bulk-like as confirmed by these experiments.

\begin{figure}
	\centering
	\includegraphics[width = 1\linewidth]{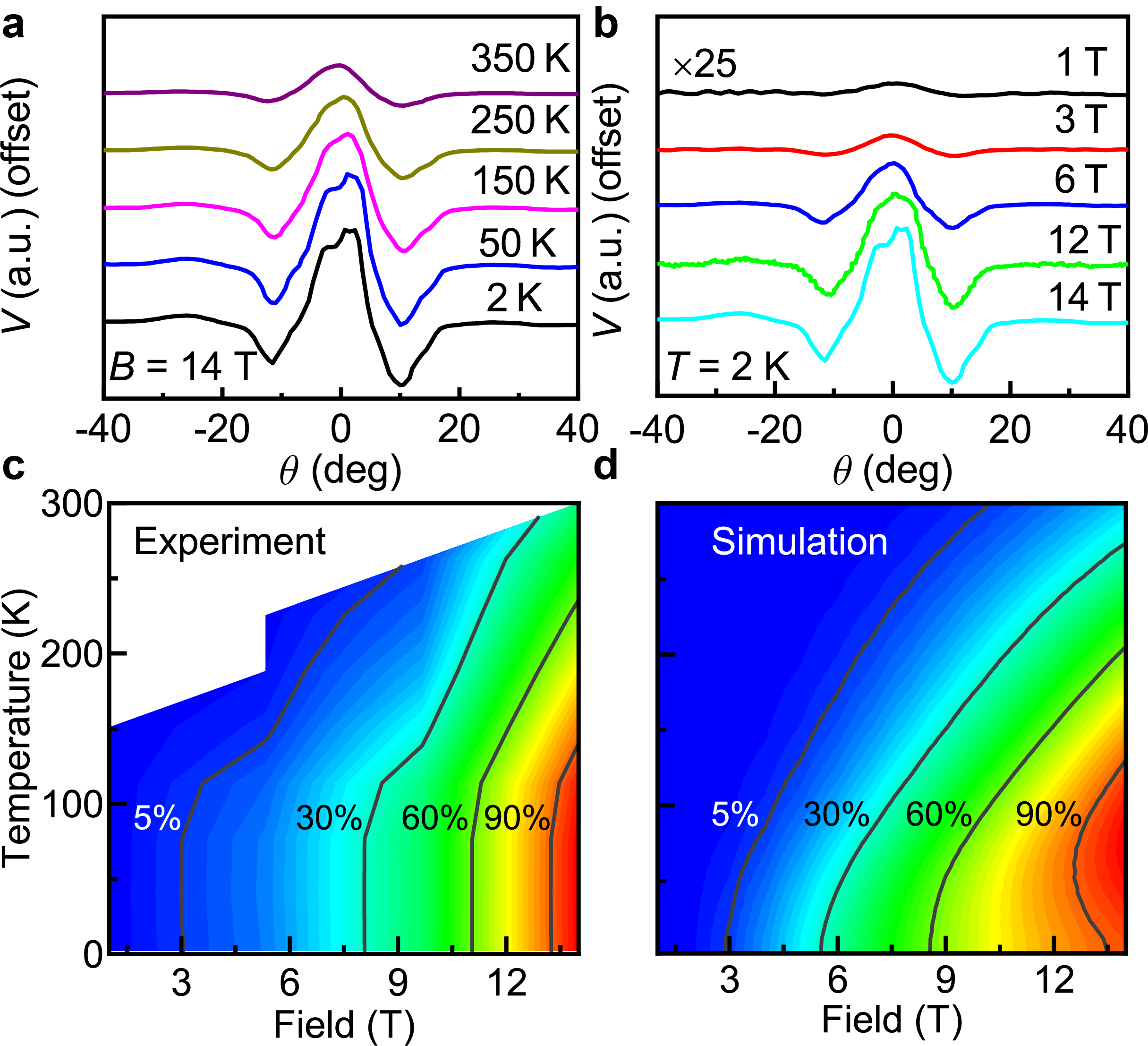}
	\caption{\textbf{a}, Temperature dependence of the bottom voltage at 14\,T. \textbf{b}, Bottom voltage at 2\,K for different magnetic fields (data at 1\,T has been multiplied by 25 and the curves in both (a) and (b) have been offset for clarity). \textbf{c}, experimental data and \textbf{d}, simulation of current collimation strength $\Delta$ depending on magnetic field and temperature. $\Delta$ is estimated by the difference between the minimum and maximum value of the signal. The color encodes the relative size of $\Delta$ with respect to the highest measured value at $\Delta$(2\,K,14\,T).}
\label{Fig3}
\end{figure}

Both the three-dimensional nature of the sample and the out-of-plane Hall conductivity are crucial in establishing the current beams. To elucidate its role and contrast our results to two-dimensional systems, we simulate the electric potential with the same anisotropy on the diagonal conductivity yet with an artificially suppressed Hall effect, $\sigma_{\rm xy}\,=\,0$ (Fig.4). This would physically resemble a two-dimensional system with a strong in-plane anisotropy induced by an in-plane magnetic field, for example due to spin-scattering. While the pronounced anisotropy still results in non-local current patterns spanning the entire device, the characteristic flow pattern is distinctly different. Instead of two well-defined beams terminating at the contacts, a simple potential gradient between the electrodes is established. Such a situation leads to a well distinguishable voltage signature, in particular the voltage remains strictly positive and the minima are absent. Without the out-of-plane Lorentz force preserving the helical motion, the beams spread out into a broad diffusive pattern. This exemplifies that the three-dimensional motion of charge carriers is key to the behavior observed in these devices, leading to transport signatures unobtainable in 2D systems.


\begin{figure}
	\centering
	\includegraphics[width = 1\linewidth]{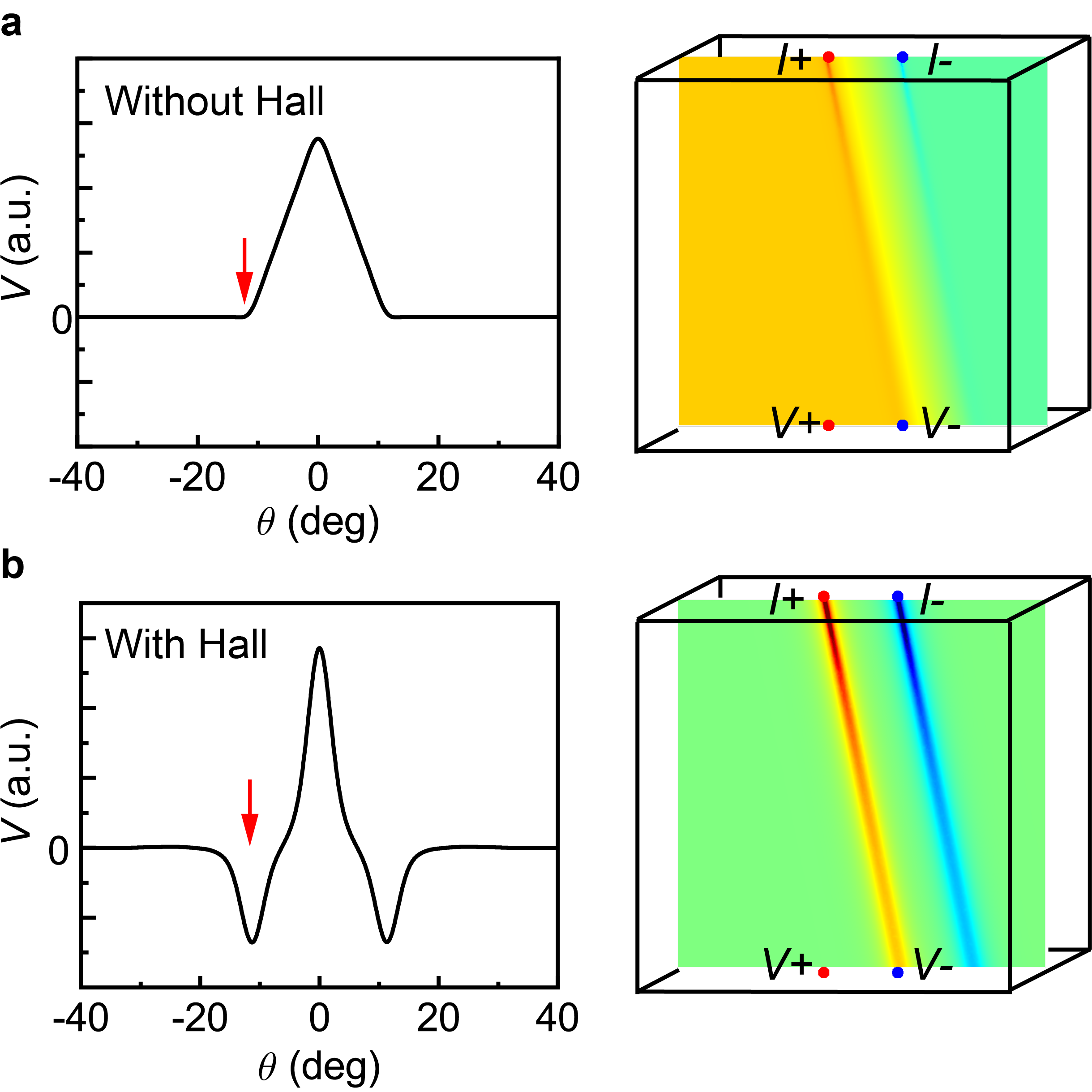}
	\caption{
	\textbf{a}, non-local voltage simulations using the same conductivity anisotropy, yet artificially setting $\sigma_{\rm xy}\,=\,0$. \textbf{b}, compare to the simulation using including a realistic $\sigma_{\rm xy}$ as in all other simulations. Both cases lead to positive non-local voltage, yet the appearance of the pronounced minimum is a direct consequence of the 3D nature and requires a strong $\sigma_{\rm xy}$. The color scale is offset between panels for clarity.}
\label{Fig4}
\end{figure}

To demonstrate the feasibility of more complex architectures and functionality, we turn to multi-contact devices shown in Fig.5. Here two rows of 4 contacts, $\sim$ \SI{1}{\micro\metre} wide spaced $\sim$ \SI{2}{\micro\metre} apart, are separated by a $\sim$ \SI{10}{\micro\metre} channel. A current sourced between two contacts in the top row leads to the same generation of current beams as before. By rotating the magnetic field in the plane, the signal propagation to the contacts in the bottom row can be selected. For fields at $17^{\circ}$, the beams directly hit the voltage pair (c,d) while no signal is observed at the opposite pair (e,f). The current jet is fully determined by the shape of the injection points, allowing additional freedom to design arbitrary desired electrical responses. In particular, the peak amplitude, angle position and asymmetry are sensitively determined by the contact geometry. Here, the contact width and separation was varied throughout the device. While differences in contact width lead to asymmetry, differences in contact separation between injection and detection pairs reduce the peak amplitude as expected.

\begin{figure}
	\centering
	\includegraphics[width = 1\linewidth]{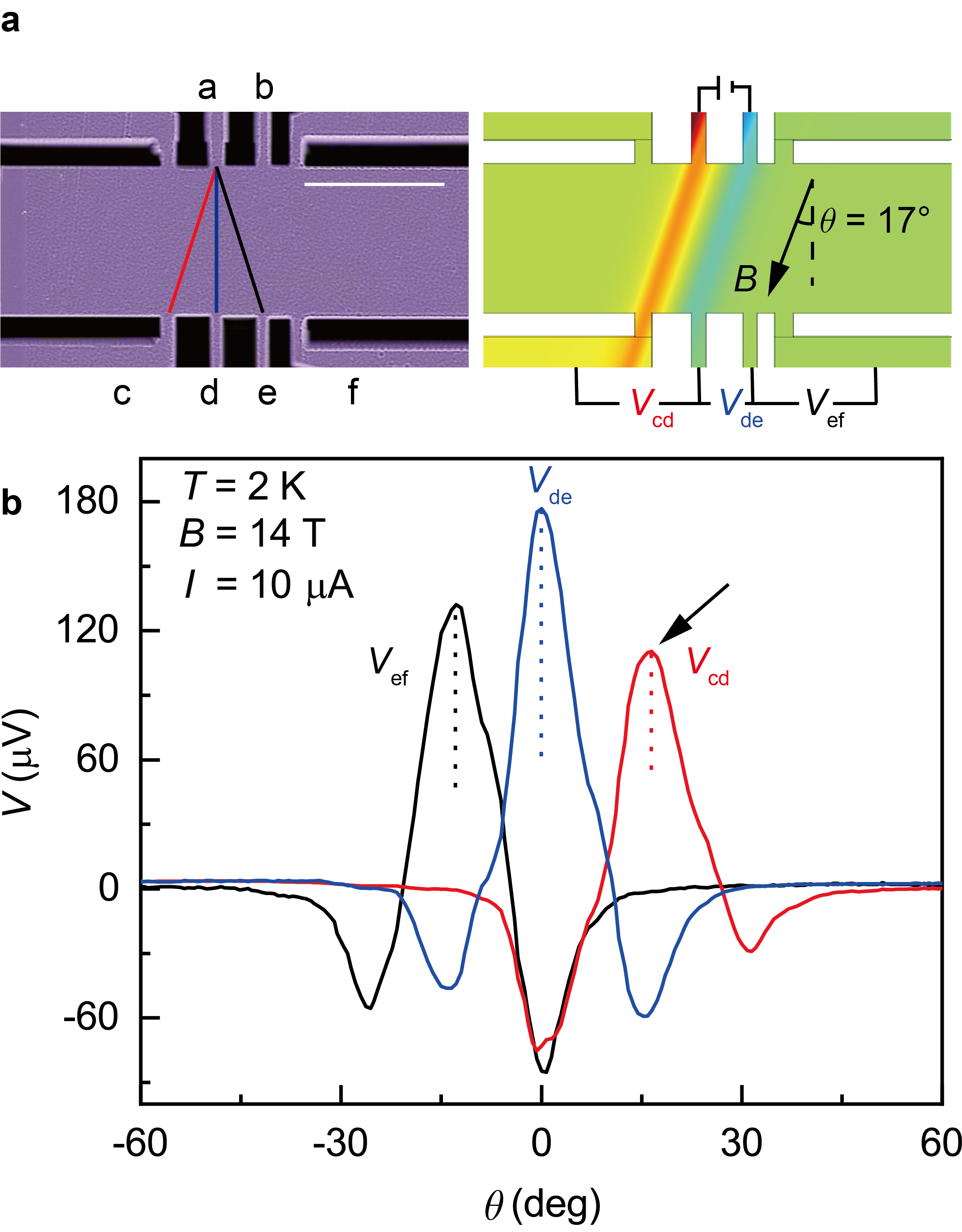}
	\caption{\textbf{a},  False-colour electron micrograph of a Cd$_{3}$As$_{2}$ (purple) multi-channel device. The scale bar is \SI{10}{\micro\metre} in width. The colored solid lines are schematic illustration of the electron beams when the magnetic field is along different angles shown by the dashed lines in (\textbf{b}). The current is sourced between the contacts $a$ and $b$, and all pairs of bottom voltages are measured. The arrow points the position when the magnetic field is at $17^{\circ}$ as shown in the simulation of (\textbf{a}).}
\label{Fig5}
\end{figure}

The geometry is designed such that the distance between adjacent nozzles matches with the geometric criterion of the minimum position $\theta_{\rm g}$. This leads to interesting signals at $0^{\circ}$: While the beams directly hit the opposite nozzles leading to a large, positive $V_{\rm de}$, the other pairs show the maximal negative voltage. As is, the simple device acts as a basic solid-state demultiplexer or a digital magnetic field direction sensor, capable of operation in extreme magnetic field and low temperatures where common semi-conductor sensors such as Hall bars become non-linear due to quantum effects.


These experiments demonstrate narrow electron jets in solids exploiting the large Hall conductivity and magneto-resistance of low-carrier density semi-metals. It will be interesting to explore further how the field-tunable anisotropy and the highly non-local signal propagation may be exploited in novel applications. For example, neuromorphic computation schemes depend on non-local signal propagation to simulate neuronal behavior which could be, in principle, implemented through current jetting devices\cite{Indiveri2011}. Device operation without the need for external magnetic fields could be implemented through magnetic thin films and spintronics technology. While the complexity of three dimensions has been treated as a nuisance in the past, we demonstrate that embracing it might eventually lead to a new class of applications.


\section{Methods}

\textbf{Focused Ion Beam sample preparation}: Starting from a millimeter-sized flux-grown single crystal, the orientation of the crystal is determined by X-ray diffraction such that the crystalline $c$ axis is perpendicular to the main surface of the slab cut from the crystal. The devices S1-S6 were machined using both a FEI Helios G3 FIB using Ga-ions and a FEI Helios PFIB using Xe-ions and the multi-channel device was fabricated in a ZEISS Crossbeam FIB. A slab or ``lamella" ($\sim$ \SI{150}{\micro\metre} $\times$ \SI{50}{\micro\metre} $\times$ \SI{3}{\micro\metre}) was cut at 30\,kV and a current of 60\,nA for coarse and 4\,nA for finer milling (PFIB). The slab then was transferred from the crystal to a sapphire substrate ex-situ using a micro-manipulator under an optical microscope. The samples were fixed using two-component araldite rapid epoxy. Low ohmic contacts were achieved by gold deposition [A) argon etching at 200\,V for 5\,min; B) 5\,nm Ti; C) 200\,nm Au]. In a second FIB step, the gold film was partially removed using the FIB at 2\,kV and a current of 1.7\,nA (Ga), clearing the region of interest on the device from the gold top layer. Next, the final structure was cut from the slab at 30\,kV and a current of 0.79\,nA for coarse and 80\,pA for fine milling (Ga). Lastly, the contact of the device was polished at an current of 7.7\,pA. Six samples of varying dimensions were prepared following this recipe. All sample dimensions were measured using a scanning electron microscope and are listed in Supplementary Table 1.

\textbf{Finite-element simulations}: Simulations of the potential distribution are based on solving the stationary Laplace equation and were performed using the finite element software package Comsol Multiphysics 5.3a. 3D models of the devices were used as boundary conditions, and the current and voltage distributions have been calculated self-consistently. The conductivity $\sigma$ when the magnetic field is along $\mathbf{z}$ direction is determined based on the Drude model\cite{pippard2009magnetoresistance}:

$\sigma=\begin{pmatrix}
\frac{\sigma _{0}}{1+(\mu B)^{2} } &\frac{\sigma _{0}\cdot (\mu B)}{1+(\mu B)^{2} }  & 0\\ 
\frac{\sigma _{0}\cdot (\mu B)}{1+(\mu B)^{2} } & \frac{\sigma _{0}}{1+(\mu B)^{2} } & 0\\ 
0&0  & \sigma _{0}
\end{pmatrix}$

$\sigma _{0}$ is the conductivity of the material in zero magnetic field. $\mu$ is the carrier mobility and assumed to be field-independent, $B$ is the magnetic field. 
The in-plane field rotation can be simply introduced by a matrix transformation of the tilted conductivity tensor into the reference frame of the device, $\sigma{}'=R_{\rm y} \cdot \sigma \cdot R_{\rm y}^{-1}$ . $R_{\rm y}$ is the 2D rotation matrix 

$R_{\rm y}=\begin{pmatrix}
\rm cos(\theta)  &0  & \rm sin(\theta) \\
0 & 1 & 0\\
-\rm sin(\theta)  &0  & \rm cos(\theta)
\end{pmatrix}$

The reported voltage at a contact was obtained by averaging the electric potential on the surface of the voltage contact.

\textbf{Transport Measurement}: The transport measurements are performed in a Physical Property Measurement System (PPMS) with a 14\,T superconducting magnet. In the measurement, a typical ac excitation current $I_{\rm ac}$\,=\,\SI{10}{\micro A} at 177.77\,Hz is applied to the two top contacts and the voltage is measured between the bottom contacts. Angular-dependent transport measurements are facilitated by the sample rotator of the PPMS. The sample was rotated in the ab plane from $-90^{\circ}$ to $90^{\circ}$ and data points were taken continuously while rotating.


\bibliography{paper}

\end{document}